# Enhancement of Sensitivity-Bandwidth Product of Interferometric GW Detectors using White Light Cavities

M. Salit[1*] and M.S. Shahriar[1,2]

[1]Department of Physics and Astronomy
[2]Department of Electrical Engineering and Computer Science
Northwestern University, 2145 Sheridan Rd, Evanston, IL. 60208, USA
[*]Corresponding author: m-salit@u.northwestern.edu

**Abstract:** The effect of gravitational waves (GWs) has been observed indirectly, by monitoring the change in the orbital frequency of neutron stars in a binary system as they lose energy via gravitational radiation. However, GWs have not yet been observed directly. The initial LIGO apparatus has not yet observed GWs. The Advanced LIGO (AdLIGO) will use a combination of improved techniques in order to increase the sensitivity. Along with power recycling and a higher power laser source, the AdLIGO will employ signal recycling (SR). While SR would increase sensitivity, it would also reduce the bandwidth significantly. Previously, we and others have investigated, theoretically and experimentally, the feasibility of using a White Light Cavity (WLC) to circumvent this constraint. However, in the previous work, it was not clear how one would incorporate the white light cavity effect. Here, we first develop a general model for Michelson-Interferometer based GW detectors that can be easily adapted to include the effects of incorporating a WLC into the design. We then describe a concrete design of a WLC constructed as a compound mirror, to replace the signal recycling mirror. This design is simple, robust, completely non-invasive, and can be added to the AdLIGO system without changing any other optical elements. We show a choice of parameters for which the signal sensitivity as well as the bandwidth are enhanced significantly over what is planned for the AdLIGO, covering the entire spectrum of interest for gravitational waves.

**PACS Number(s):** 42.50.Ct, 42.50.Gy, 42.60.Da

## 1. Introduction

Astronomers and optical scientists have often worked together to do astronomy, and gravitational wave (GW) astronomy will not be an exception. GW detectors are not optical devices in the sense that telescopes are, but the most promising of them use interferometers to sense gravitational radiation by virtue of its effect on laser light here on earth [1]. Here we deal with the optics of laser interferometric GW detectors. We analyze the frequency response and sensitivity for several potential designs, including a proposed modification that uses a White Light Cavity (WLC) to enhance the sensitivity-bandwidth product. We previously demonstrated a WLC experimentally in rubidium [2], and have also explored photorefractive crystals as a potential medium for adapting the technique for use at the working wavelength of LIGO.[3,4] We review the theory of the WLC and show mathematically the advantages it can offer for LIGO-type GW detectors.

When light travels through a region of space over which a GW is also propagating, the latter causes a periodic variation in the phase of the light field.[5] Mathematically, light with this kind of phase modulation may be described as a sum of plane waves of different frequencies. The largest frequency component is the carrier, which is just the frequency of the light when the modulation amplitude is set to zero. The next largest are the two first order sidebands: a Plus-Sideband at the carrier plus the modulation frequency, and a Minus-Sideband at the carrier minus the modulation frequency.[6] Higher order sideband frequencies exist; however, when the modulation is small, as in the case of GWs, their amplitudes are negligible. The problem of detecting GWs may be reduced to the problem of detecting these sideband frequencies.

The difficulty lies in the fact that the amplitudes of these sideband frequency components are very small, and that they are expected to be separated generally by less than a few kilohertz, and in some cases by only tens or hundreds of hertz, from the carrier frequency. These sidebands, then, cannot be separated out from the carrier by means of the usual techniques for filtering light. Prisms and diffraction gratings will not resolve such tiny frequency differences, and even Fabry-Perot cavity filters are less than ideal for this purpose, as they would have to have linewidths down to tens of hertz and very high transmittivity on resonance, so as not to further attenuate the already weak sidebands.

Fortunately we can take advantage of a very convenient property of gravitational radiation: the fact that the modulations it causes along one axis are exactly out of phase with the modulations along a perpendicular axis.[7] We can therefore use an interferometer to separate out the carrier and the sidebands. Both Michelson and Sagnac interferometers have been proposed for this purpose. We discuss the Sagnac case in reference 8. Here, we will discuss GW detectors that are variations on the Michelson interferometer.

If we arrange the arms of the interferometer along the x and y axes, and the path lengths are chosen correctly, then at one port the carrier light from the x-axis will exactly cancel the carrier light from the y-axis so that we get no carrier frequency light out. The interferometer is on a dark fringe for the carrier. The sidebands, however, having been created by phase modulations with opposite signs, will interfere constructively at this same port.[6] This means that we can have only the sideband light exiting one port of a Michelson interferometer under the dark fringe condition. Detecting light at that port, in theory, indicates the presence of a GW.

In practice the situation is more complicated. Most of the time light at this dark port only indicates vibrations in the interferometer mirrors or other sources of noise. A great deal of work has been done to minimize noise and to lock the interferometer on a dark fringe condition,[6] but we would also like to maximize the amplitude of the sideband light falling on the detector. One way of doing that, due to the nature of GWs, is to make the arms of the interferometer very long [1].

Another approach involves the use of optical cavities within the interferometer. If the resonance linewidth of the cavities used is too small, however, then our attempts to use them to enhance the sensitivity of the GW detector will also entail narrowing its linewidth. For this reason, the ideal detector may use a White Light Cavity (WLC), to get the benefits of cavity enhancement described below, without correspondingly narrowing the linewidth of the detector. A WLC is a cavity that resonates over a broader range of frequencies than what its length and finesse would ordinarily entail. The basic theory behind the WLC is discussed in Section 4 below, and explored in greater detail in references 9,10,11 and 12.

The remainder of this paper is organized as follows. Section 2 discusses several GW detector designs that have been proposed, starting from the basic Michelson interferometer configuration. Section 3 gives a general derivation of the frequency response of devices of this type, including those described in the preceding section. Section 4 discusses the effect of a dispersive medium on that frequency response, and in particular, the effect incorporating WLCs into the design. We conclude in Section 5 with a summary of our results.

**2. Variations on the Michelson Interferometer**

There are a variety of ways to use cavities to improve the response of the Michelson-interferometer based GW detector. One of the simplest is to add additional mirrors in each arm of the interferometer so as to turn each arm into a Fabry-Perot cavity, as illustrated in Figure 1, below.

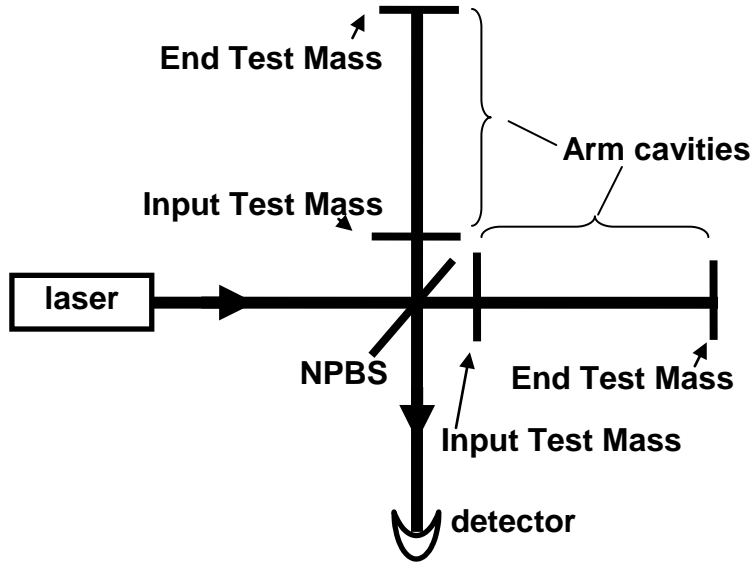

**Figure 1:** *Michelson interferometer with arm cavities.*

Sideband light is produced from the carrier on each pass as it bounces around the arm cavities. However, though the effect is similar to the use of longer cavity arms, we cannot simply model this as a system with longer effective lengths for the arms. We must take into account the inference effects of multiple bounces within these arm cavities.

We might choose to make the arm cavities resonant for the carrier frequency, for instance. This would allow us to increase the amplitude of the carrier frequency field in the arms by a potentially large factor. Since the sideband field is proportional to the carrier field, the amount of sideband light produced in the arms would then be increased by this same factor. However, the sideband light itself would also undergo multiple reflections within the arm cavities. If the frequency separation between one of the sidebands and the carrier were greater than the resonance linewidth of the cavity, then the multiple reflections of this sideband would interfere destructively. The same conclusion would apply to the other sideband as well, and the signal at the output would be small. Similarly, we might tune the arm cavities to resonate the sidebands, but the carrier would then interfere destructively inside the cavities, making the net signal small.

One way to avoid this trade-off is to place the cavity input mirrors outside the arms of the interferometer, as illustrated in Figure 2, below:

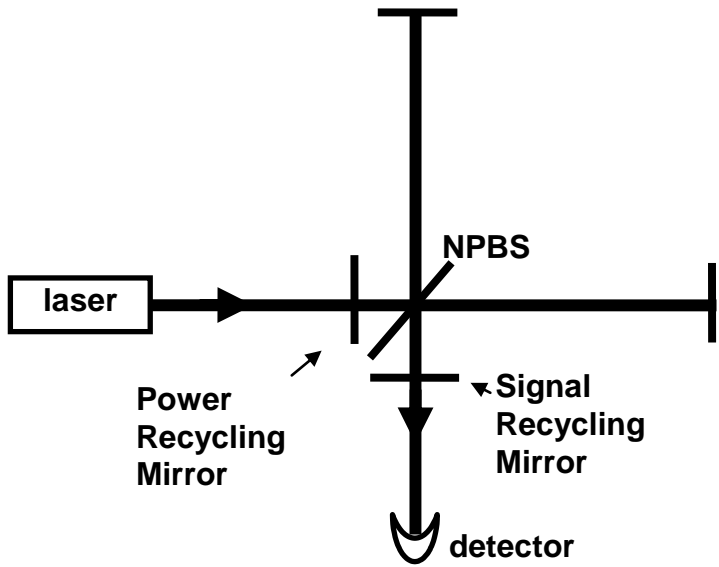

**Figure 2:** *Michelson Interferometer with dual recycling*

In this configuration, assuming the interferometer is held on the dark fringe condition, the carrier light will only be incident on the mirror labeled Power Recycling Mirror (PRM). It will undergo multiple reflections inside both arms, as if there had been an input mirror in each. Similarly, the sideband frequency light will be reflected back into the arms by the Signal Recycling Mirror (SRM). This Dual Recycling arrangement allows both the carrier and one or both of the sidebands to resonate, within separate but overlapping optical cavities. The disadvantage of this scheme is that the beamsplitter is inside the optical cavity in which the carrier resonates. The current design for Advanced LIGO proposes a circulating power in the arms of 800kW [13]. This amount of power causes thermal distortion and noise on the beamsplitter.

A third option is to combine these two designs, as illustrated in Fig. 3, below.

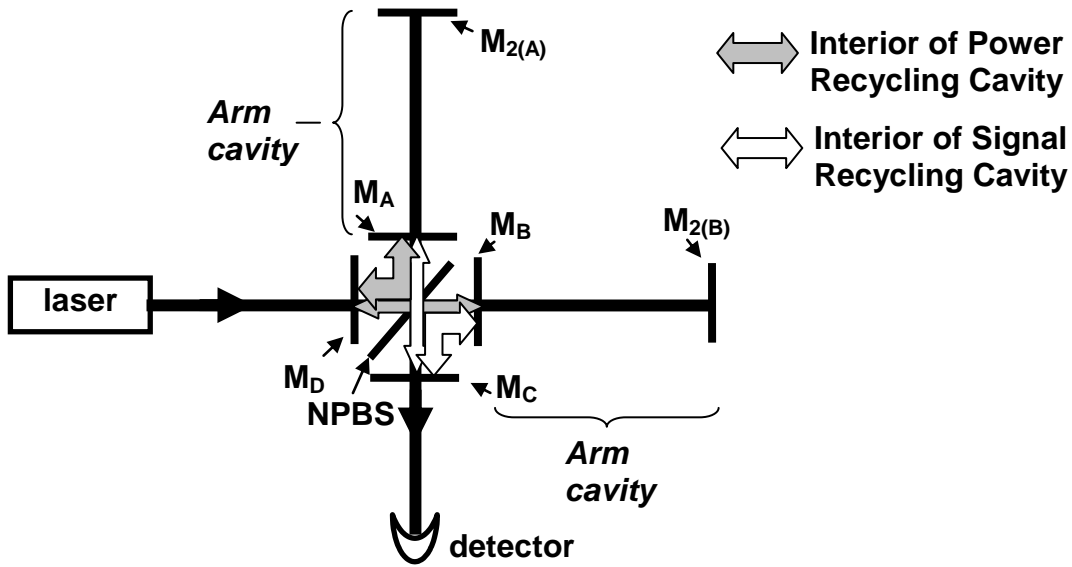

**Figure 3:** *Michelson Interferometer with dual recycling and arm cavities*

This system, though it comprises many overlapping compound cavities, is not much more difficult to analyze than the simpler version from Fig. 1, under certain conditions. If the two arm cavities are completely identical, with the reflectivity and position relative to the beamsplitter for the mirrors $M_A$ and $M_{2(A)}$ being exactly the same as those for mirrors $M_B$ and $M_{2(B)}$, then we may cease to distinguish between the end test masses $M_{2(A)}$ and $M_{2(B)}$, and refer simply to $M_2$.

Likewise, since we have assumed $M_A$ and $M_B$ are identical, we might simply refer to $M_{AB}$ to indicate either one of these input test mass mirrors. Carrier light that is incident on either one from the arms will then travel toward $M_D$, so long as the interferometer is locked on a dark fringe. Having reflected off of $M_D$ it will then travel back to one of the mirrors $M_{AB}$, and then back toward $M_D$ again, so that a cavity is formed. We will refer to this cavity as the Power Recycling Cavity (PRC).

The sideband light, likewise, travels from $M_{AB}$ to $M_C$ and back again, so that the sidebands experience a different cavity than the carrier. We will refer to this cavity as the Signal Recycling Cavity (SRC).

In general, any Fabry-Perot cavity may be treated, from the outside, as a mirror that has a frequency dependent reflectivity. Therefore, we treat the PRC, comprising $M_{AB}$ and $M_D$, as a single compound mirror $M_{1CAR}$, because it is the compound mirror which reflects the carrier back into the arms. Likewise, we treat the SRC, comprising $M_{AB}$ and $M_C$, as a single compound mirror $M_{1SB}$, because it is the compound mirror which reflects the sidebands back into the arms.

The total system may then be modeled as a single Fabry-Perot cavity, with one mirror $M_2$ having a reflectivity equal to that of the end test masses $M_{2A}$ and $M_{2B}$, and one mirror $M_1$, whose reflectivity is frequency dependent, equal to that of the compound mirror $M_{1CAR}$ for carrier frequency light, and equal to that of the compound mirror $M_{1SB}$ for sideband frequency light. The length of this effective cavity is equal to the distance between $M_A$ and $M_{2A}$ (or equivalently, between $M_B$ and $M_{2B}$). This model is illustrated in Fig. 4 below.

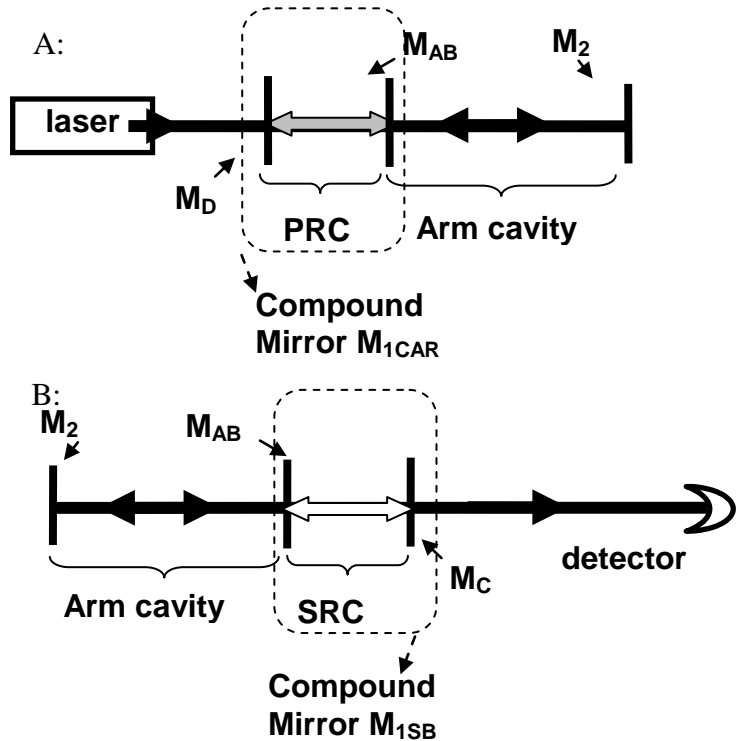

**Figure 4:** *A) Effective path for carrier through system illustrated in Fig. 3 B) Effective path for GW sideband through system illustrated in Fig. 3*

We can now choose the length of the arm cavities to resonate one of the sidebands, the carrier, or both. However, we have the choice, thanks to the fact that we now have tunable compound mirrors, to make the finesse different for the sidebands than for the carrier. The next section analyzes the behavior of this system in mathematical detail. In this section, however, we first summarize the results qualitatively.

There are two different modes of operation for this device[14]. In both, we choose the reflectivity of $M_{1CAR}$ to be high (by tuning the PRC length and making the reflectivity of $M_D$ high) and choose the length of the arms so that the carrier is resonant in the arm cavities. Since the sidebands necessarily have different wavelengths than the carrier, they will be off resonant in the arm cavities. Therefore, if we do not want the output signal to be nullified by the destructive interference of the sideband light in the arms, we must either lower the reflectivity of the compound output coupler $M_{1SB}$(by tuning the SRC), thus broadening the arm cavity linewidth enough to allow the sidebands to survive, or try to tune the phase the light picks up on reflection from this compound mirror to bring the sideband light back to resonance in the arms.

The idea of lowering the finesse of the arm cavities for the sidebands by using a near-resonant cavity as an output coupler is called *Resonant Sideband Extraction* (RSE) [15]. The SRC is much shorter than the arm cavities, on the order of 50 m[16] as opposed to 4000 m, and has a correspondingly broader linewidth – at least 30 kHz even with the highest reflectivity mirror choices we have used in the models in the next section. For realistic choices of mirror reflectivities, if the length of the SRC is chosen such that the carrier frequency would resonate in it, we find GW sidebands within the spectrum of interest are also transmitted effectively. This mode of operation, with the length of the SRC chosen to resonate carrier frequency light, is conventionally known as *tuned* or *symmetrically tuned* mode. It is symmetric in the sense that sidebands spaced equally above and below the carrier frequency transmit equally. The technique allows us to build up the carrier field in the cavities, increasing the amplitude of our output signal, without destroying the sideband fields. The response is peaked at zero gravitational wave frequency but is relatively broadband, as we will see.

In the other mode, we attempt to adjust the phase of the reflectivity of the compound output coupler such that at least one sideband frequency is resonant in the arm cavities. Changing this reflectivity has no effect on the resonance of the carrier; for which only the PRC reflectivity is relevant. Again, because the SRC linewidth is broad enough to encompass the range of GW sidebands of interest, the phase of this reflectivity is relatively uniform over the spectrum of interest. In general it will exactly offset the excess phase (or the phase deficit) picked up in propagating through arm cavity only for a sideband of a particular frequency. The linewidth of this resonance depends on the magnitude of the reflectivity associated with this phase, but also on the length of the arm cavities. Making the reflectivity of the SRC higher increases the sideband field at the resonant frequency and thus the sensitivity of the detector at that frequency, but due to the great length of the arm cavities, the result is that the detector bandwidth becomes narrower than the spectrum of interest. This mode is conventionally known as *detuned* or *asymmetrically tuned* operation. Only sidebands created by a narrow range of GW frequencies, determined largely by the arm length we have chosen, are detectable in this mode. The amount of light falling on the detector is, however, higher in this narrowband mode, when the appropriate GW frequency is present, than it would be in the broadband mode for that same frequency.

Though this system offers better response in both modes than that depicted in Fig. 1, and eliminates many of the heating problems posed by that depicted in Fig.2, it still forces us to choose between high sensitivity with low bandwidth, or high bandwidth with low sensitivity. The WLC proposal that will be described in Section 4 would allow us to have at least the sensitivity of the narrowband mode over a spectrum as wide as that offered by broadband mode, so that a WLC-enhanced LIGO type interferometer might offer the best of both worlds. In fact, we will show that the sensitivity can actually be much higher than that of the narrowband mode, while allowing a bandwidth as large as the broadband mode.

## 3. General Model for Michelson-based GW Detectors

The system depicted in Fig. 3 is also a more general case of those depicted in Fig. 1 and Fig. 2. If we model it mathematically and find its response, we may recover the response for the systems depicted in Fig. 1 and Fig. 2, or for a simple Michelson interferometer with no cavities, by setting the reflectivities of the appropriate mirrors equal to zero. In this section, we develop this general model.

Meers[17] first modeled this system in 1989, but his model is not given in a form that lends itself to the analysis of the effect of a WLC on the system. Furthermore, in the course of developing the model, he makes certain assumptions about mirror reflectivities and resonance conditions which render his final expression less than general. We will essentially follow his method in deriving the frequency response of the system illustrated in Fig. 3, but we will adopt a slightly different notation and avoid assuming any particular operating condition.

As Meers did, we will denote the reflectivity of the compound mirror $M_{1CAR}$ by $R_{1C}$. This is not to be confused with the reflectivity of the mirror labeled $M_C$, which we will denote simply by $R_C$. We will denote the reflectivity of the compound mirror $M_{1SB}$ by $R_{1S}$. Whereas he uses $R_{1S}$ and $R_{1C}$ to denote only the amplitude of the reflectivity, we will allow them to be complex numbers, giving information about both the amplitude and the phase of light reflected off the SRC and PRC, respectively. We can calculate these reflectivities from basic theory of a Fabry-Perot cavity, keeping in mind that they are frequency dependent quantities wherever we use them.

The first step in the derivation is to quantify the effect of a GW on light. Let us choose our coordinates such that the effect of the GW on the metric of space-time is described by[7]

(1) $$ds^2 = dx^2(1+h\cos\omega_g t) + dy^2(1-h\cos\omega_g t) + dz^2 - c^2 dt^2$$

Along the path of a light wave, ds = 0. Let us assume we have light propagating along the x-axis. Then:

(2) $$dx^2(1+h\cos\omega_g t) = c^2 dt^2$$
$$\Rightarrow \frac{dx}{dt} \approx c(1-\frac{h}{2}\cos\omega_g t)$$

The phase the light accumulated as it travels is given by:

(3) $$\phi_x = \int_{x_1}^{x_2} k\, dx = \int_{t-\tau}^{t} k\frac{dx}{dt} dt = \int_{t-\tau}^{t} kc\left(1-\frac{h}{2}\cos\omega_g t\right) dt$$
$$\Rightarrow \boxed{\phi_x = \omega\tau - \frac{\omega h}{\omega_g}\sin\left(\frac{\omega_g \tau}{2}\right)\left(\frac{e^{i\omega_g(t-\tau/2)}+e^{-i\omega_g(t-\tau/2)}}{2}\right)}$$

The calculation for a beam traveling along the y-axis is identical, except that we use $\frac{dy}{dt} \approx c(1+\frac{h}{2}\cos\omega_g t)$.

(4) $$\Rightarrow \boxed{\phi_y = \omega\tau + \frac{\omega h}{\omega_g}\sin\left(\frac{\omega_g \tau}{2}\right)\left(\frac{e^{i\omega_g(t-\tau/2)}+e^{-i\omega_g(t-\tau/2)}}{2}\right)}$$

Of course $\omega\tau$ is the phase that the light would pick up in the absence of GWs. We define $\phi_{prop} = \omega\tau$ as the ordinary propagation phase. In our model, we assume that light traveling along one of the coordinate axes under the influence of GWs picks up a multiplication factor expressed as $e^{i\phi_x} = e^{i\phi_{prop}}e^{i\delta\phi_x} \cong e^{i\phi_{prop}}\left(1+i\delta\phi_x\right)$ (where $\delta\phi_x = \phi_x - \phi_{prop}$) or as $e^{i\phi_y} \cong e^{i\phi_{prop}}\left(1+i\delta\phi_y\right)$ (where $\delta\phi_y = \phi_y - \phi_{prop}$). By using these approximations, we are assuming that the modulation is small enough that the carrier power is effectively undepleted.

First we will consider the amplitude of the carrier field. Let us assume that a field with amplitude $E_0$ enters through $M_{1CAR}$, which has a transmittivity $T_{1C}$ and a reflectivity $R_{1C}$ at the carrier frequency. The field, after entering and reflecting off of either arm-end mirror $M_2$ (which has a reflectivity $R_2$) returns to the PRC with an amplitude

(5) $\quad E_1 = E_0 T_{1C} R_2 e^{-2ik_c L}$

where L is the length of the arm-end cavity, and $k_c$ is the carrier wavenumber.

This field now reflects off of $M_{1CAR}$, and then off of $M_2$ again, returning to the PRC now with an amplitude

(6) $\quad E_2 = E_0 T_{1C} R_2^2 R_{1C} e^{-4ik_c L}$

After each reflection thereafter the field picks up the same factor of $R_{1C} R_2 e^{-2ik_c L}$. The steady state field is the sum $\left(\sum_N E_N\right)$ over all bounces. Therefore in steady state the carrier frequency field inside is

(7) $\quad E'_{car} = E_0 T_{1C} R_2 e^{-2iNk_c L} \sum_{N=1}^{\infty} \left(R_2 R_{1C} e^{-2ik_c L}\right)^{N-1}$

Again, we have neglected the depletion of the carrier due to the modulation, in this model. Now the sideband fields being continually produced from this steady-state carrier are given, under the approximation described above, by

(8) $\quad E_{SB} = E'_{car} e^{i\omega t} e^{i\phi_{prop}} (1+i\beta\left(e^{i\omega_g(t-\tau/2)} + e^{-i\omega_g(t-\tau/2)}\right))$ where $\beta = \dfrac{h\omega}{\omega_g}\sin(\omega_g \tau/2)$

The sidebands are reflected by the SRC in figs. 3 and 4, and by the SRM in fig. 2.. Considering only the component at frequency $(\omega+\omega_g)$, we see that its initial amplitude is

(9) $\quad E_{+_1} = E'_{car} e^{i\omega t} e^{-2ik_c L} i\beta e^{i\omega_g(t-\tau/2)}$

Here we have used $\phi_{prop} = -2ik_c L$. This field reflects off the SRC and experiences a reflectivity $R_{1S}$. After another round trip the amplitude is

(10) $\quad E_{+_2} = E'_{car} e^{i\omega t} i\beta e^{i\omega_g(t-\tau/2)} R_{1S} R_2 e^{-2i\left(\frac{\omega+\omega_g}{c}\right) L_S} e^{-2ik_c L}$

Note that $(\omega+\omega_g)/c = k_+$, the wavenumber of the sideband. We have, in this expression, introduced another variable $L_S$, which is the length of the cavity in which the sidebands are propagating. In the case illustrated by figures 3 and 4, this $L_S$ is the same as $L$, equal the distance between the end test mass, $M_2$, and the input test masses, $M_{AB}$. However, in the case illustrated by figure 2, these are two distinct numbers, with $L_S$ being equal to the sum of the distance from the end test mass to the beamsplitter and that from the beamsplitter to the signal recycling mirror, and $L$ being equal to the sum of the distance from the end test mass to the beamsplitter and that from the beamsplitter to the power recycling mirror. After n passes, then, the total field is

$$(11) \quad E'_+ = \left(\sum_n E_{+_n}\right) = E'_{car} e^{i\omega t} i\beta e^{i\omega_g(t-\tau/2)} e^{-2ik_c L} \sum_{n=1}^{\infty} R_{1S}^{n-1} R_2^{n-1} e^{-2i(n-1)(\omega+\omega_g)L_S/c}$$

Doing the geometric series sums for $E'_{car}$ and $E'_+$, we find that the output field transmitted through the SRC, $E_+ = E'_+ T_{1S}$ is given by

$$(12) \quad \frac{E_+}{E_o e^{i\omega t}} = \frac{T_{1S} T_{1C} R_2}{1 - R_2 R_{1C} e^{-2ik_c L}} \frac{i(h\omega/\omega_g)\sin(\omega_g \tau/2) e^{-2ik_c L} e^{i\omega_g(t-\tau/2)} e^{-2ik_c L}}{1 - R_{1S} R_2 e^{-2i(\omega+\omega_g)L_S/c}}$$

The notation here is slightly different from that used by Meers[17], but the result agrees with his provided we define $2L/c = \tau$, $\delta_C = (-2\omega L/c) \bmod 2\pi$ and $\delta_S = (-2\omega L_S/c) \bmod 2\pi$. By leaving the expression in terms of the separate wavenumbers of the sidebands and carrier, however, we leave ourselves the option of easily including dispersive effects in this calculation at the next stage.

For the Minus-Sideband, the expression is the same, except with $\omega_g \to -\omega_g$, and with $R_{1S}$ potentially taking on a different value, since it is a frequency dependent reflectivity. These amplitudes do not tell us the frequency response of our device directly, however. In practice, the sidebands are detected by allowing a small amount of carrier frequency light to leak through, and detecting the beat signal. To find the total response of the interferometer we need to calculate the amplitude of that beat signal:

$$(13) \quad \delta I = E_L E_+^* + E_+ E_L^* + E_L E_-^* + E_- E_L^*.$$

Here $E_L$ is the carrier frequency field with which we are mixing our sidebands:

$$(14) \quad E_L = (A/E_0) e^{i(\omega t + \phi)}$$

In order to do this sum, it is convenient to change our notation slightly. Let $R_{1C} = r_{1C} e^{\phi_{r1C}}$, and let $R_{1S_+} = r_{1S_+} e^{\phi_{r1S_+}}$ be the reflectivity of the SRC at the Plus-Sideband frequency, while $R_{1S_-} = r_{1S_-} e^{\phi_{r1S_-}}$ is the reflectivity of the SRC at the Minus-Sideband frequency. In general, lower case letters for the reflectivity or transmittivity will now be used to denote the magnitude only. We also choose to insert a couple of multiplicative factors equal to one, marked with square brackets. With this convention the equation above may be rewritten as:

**(15)**

$$\frac{E_+}{E_o e^{i\omega t}} = \frac{t_{1S_+} e^{\phi_{t1S_+}} t_{1C} e^{i\phi_{t1C}} r_2 i h\omega \sin(\omega_g \tau/2) e^{i\omega_g (t-\tau/2)}}{\omega_g} \frac{e^{-2ik_cL}\left[e^{i\phi_{r1C}} e^{-i\phi_{r1C}}\right]}{\left(1 - r_2 r_{1C} e^{i\phi_{r1C}} e^{-2ik_cL}\right)} \frac{e^{-2ik_cL}\left[e^{i\phi_{r1S_+}} e^{-i\phi_{r1S_+}}\right]}{\left(1 - r_2 r_{1S_+} e^{i\phi_{r1S_+}} e^{-2ik_+L_S/c}\right)} \left[e^{-2ik_+L_S} e^{2ik_+L_S}\right]$$

These additional factors allow us to make use of the identity $\frac{e^{i\phi}}{1 - \rho_1 \rho_2 e^{i\phi}} = \frac{e^{i\phi} - \rho_1 \rho_2}{(1-\rho_1\rho_2)^2 (1 + F' \sin^2(\phi/2))}$, where $F' = \frac{4\rho_1\rho_2}{(1-\rho_1\rho_2)^2}$, to write the output in terms of a cavity finesse. We will use $F'_C$ for the finesse of the cavity as experienced by the carrier frequency light, $F'_{S_+}$ for the Plus-Sideband and $F'_{S_-}$ for the Minus-Sideband.

We now have

**(16)**

$$\frac{E_+}{E_o e^{i\omega t}} = \frac{t_{1S_+} t_{1C} r_2 i h\omega \sin(\omega_g \tau/2)}{\omega_g} \left(\frac{e^{-2ik_cL + i\phi_{rC}} - r_2 r_{1C}}{(1-r_2 r_{1C})^2 \left(1 + F'_C \sin^2\left(\frac{-2k_cL + \phi_{r1C}}{2}\right)\right)}\right) \left(\frac{e^{-2ik_+L + i\phi_{rS_+}} - r_2 r_{1S_+}}{(1-r_2 r_{1S_+})^2 \left(1 + F'_{S_+} \sin^2\left(\frac{-2k_+L_S + \phi_{r1S_+}}{2}\right)\right)}\right)$$

$$\times e^{i\omega_g (t-\tau/2)} e^{\phi_{t1S_+}} e^{i\phi_{t1C}} e^{-i\phi_{r1C}} e^{-2ik_cL} e^{-i\phi_{r1S_+}} e^{2ik_+L_S}$$

We would like to separate out the part of this expression that represents the sideband resonance in arms. To this end, we define

**(17)** $$\xi_\pm = \frac{t_{1C} t_{1S_\pm} r_2 h\omega \sin(\omega_g \tau/2)}{\omega_g (1-r_2 r_{1C})^2 (1-r_2 r_{1S_\pm})^2}$$

This contains all of the scaling information which is independent of the length of the arms. And we let

**(18)** $$B e^{i\phi_B} = \frac{e^{-2ik_cL + i\phi_{r1C}} - r_2 r_{1C}}{1 + F'_C \sin^2(k_c L - \phi_{r1C}/2)}$$

Now $B$ and $\phi_B$ carry the information about the magntitude and phase of the carrier field in the arm cavities. With this notation,

**(19)** $$\frac{E_+}{E_0 e^{i\omega t}} = i\xi_+ B \left(\frac{1 - r_2 r_{1S_+} e^{2ik_+L - i\phi_{r1S_+}}}{1 + F'_{S_+} \sin^2(k_+ L_S - \phi_{r1s_+}/2)}\right) e^{i\omega_g (t-\tau/2)} e^{i\phi_{eff}} e^{i\phi_{t1S_+}} e^{i\phi_B}$$

where $\phi_{eff} = \phi_{t1C} - \phi_{r1C} - 2ik_c L$

With this expression and some trigonometric identities, it is now relatively straightforward to calculate the total response of our device. In keeping track of the phase of the carrier, it proves convenient to define $\phi_{net} = (\phi_{eff} - \phi + \phi_B)$. We also replace $\tau$ with $2L/c$ at this point so as to make all length dependence explicit.

We find that

(20)

$$E_+ E_L^* + E_+^* E_L = \frac{-2AB\xi_+}{1+F_{S_+}' \sin^2(k_+ L_S - \phi_{r1S_+}/2)}$$
$$\times \left[ \sin\left(\omega_g\left(t-\frac{L}{c}\right) + \phi_{t1S_+} + \phi_{net}\right) - r_2 r_{1S_+} \sin\left(\omega_g\left(t-\frac{L}{c}\right) + \phi_{t1S_+} + \phi_{net} + 2k_+ L_S - \phi_{r1S_+}\right) \right]$$

Finally, we choose

(21) $\quad \phi_C = \phi_{net} - \pi/2 = \phi_{t1C} - \phi_{r1C} - 2k_c L - \phi + \phi_B - \pi/2$

This variable keeps track of the total phase of the carrier, and the term $\pi/2$ allows us turn our sine functions into cosine functions. Note that the unsubscripted $\phi$ comes from assuming our sidebands are beating with a carrier frequency field of the form $E_L = (A/E_0)e^{i(\omega t + \phi)}$. We will assume that this phase is controllable, and that we can always choose it so that the output is optimum. The signal from our device is then

(22)

$$\delta I = E_L E_+^* + E_+ E_L^* + E_L E_-^* + E_- E_L^*$$
$$= 2AB\left[\left(\frac{\xi_+ \cos(\omega_g(t-L/c))}{1+F_{S_+}' \sin^2(k_+ L_S - \phi_{r1S_+}/2)}\right)\left(r_2 r_{1S_+} \cos(2k_+ L - \phi_{r1S_+} + \phi_{t1S_+} + \phi_C) - \cos(t_{1S_+} + \phi_C)\right)\right.$$
$$+ \left(\frac{\xi_+ \sin(\omega_g(t-L/c))}{1+F_{S_+}' \sin^2(k_+ L_S - \phi_{r1S_+}/2)}\right)\left(-r_2 r_{1S_+} \sin(2k_+ L - \phi_{r1S_+} + \phi_{t1S_+} + \phi_C) + \sin(t_{1S_+} + \phi_C)\right)$$
$$+ \left(\frac{\xi_- \cos(\omega_g(t-L/c))}{1+F_{S_-}' \sin^2(k_- L_S - \phi_{r1S_-}/2)}\right)\left(r_2 r_{1S_-} \cos(2k_- L - \phi_{r1S_-} + \phi_{t1S_-} + \phi_C) - \cos(t_{1S_+} - \phi_C)\right)$$
$$+ \left.\left(\frac{\xi_- \sin(\omega_g(t-L/c))}{1+F_{S_-}' \sin^2(k_- L_S - \phi_{r1S_-}/2)}\right)\left(r_2 r_{1S_-} \sin(2k_- L - \phi_{r1S_-} + \phi_{t1S_-} + \phi_C) - \sin(t_{1S_-} + \phi_C)\right)\right]$$
$$\equiv P\cos(\omega_g(t-L/c)) + Q\sin(\omega_g(t-L/c))$$

To find the magnitude of this signal, then, we have only to add the amplitudes of the sine and cosine terms in quadrature.

**(23)**   $|\delta I| = \sqrt{P^2 + Q^2}$

This rather complicated expression gives the full response of the system illustrated in figure 3, if we set $L_S = L$. In the limit where $r_{1S_+} = r_{1S_-} = r_{1C}$, this also gives the response of the simpler system illustrated in figure 1. Finally, this expression can give us the response of the system illustrated in Fig. 2 as well, where $L_S \neq L$, $r_{1S_\pm}$ is equal to the reflectivity of the SRM, and, $r_{1C}$ is equal to that of the PRM.

A GW detector, in the configuration illustrated in Fig. 3 and described by the above equation, has two basic modes of operation, as previously discussed. In the narrowband mode, the SRC is tuned to be far off resonance for the sidebands, so that the reflectivity of the SRC is high, and therefore the finesse of the arm cavities is high for the sidebands. A length is chosen for the arm cavities so that a sideband of a corresponding wavelength will resonate. The PRC is tuned to near resonance for the carrier, so that the finesse of the arm cavity for the carrier is low enough to prevent destructive interference from reducing the carrier amplitude. In the broadband mode, the SRC is tuned to be near resonant so that its transmission is high, and its reflectivity low. The arm length is chosen so that the carrier will resonate, and the finesse of the arm cavities for the carrier is made large by tuning the PRC far off resonance.

Below, the response for the two cases, calculated using the equation above, is plotted. These graphs are to be compared with those shown in reference 15 for a similar system with different reflectivity and length parameters. We display the response both with the currently planned Advanced LIGO value for the SRM reflectivity $r_C$ and with a higher reflectivity, to illustrate the effect on the signal response. The higher reflectivity allows much larger signal responses but with much narrower bandwidths.

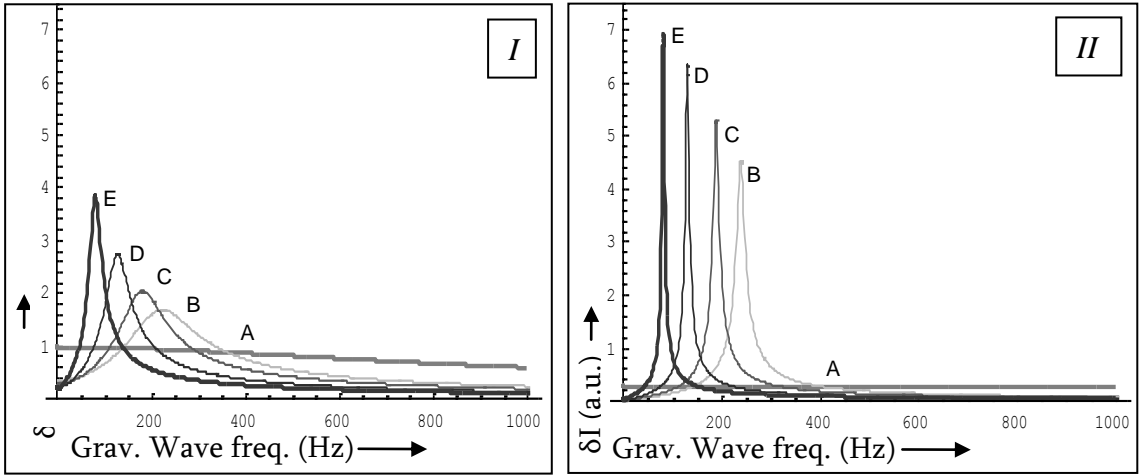

**Figure 5:** *I) Output signal as a function of gravitational frequency for a GW detector of the type illustrated in Fig. 3, using the Advanced LIGO parameters of reference 16, under different operating conditions. The tuned mode response is normalized to one at zero frequency II) The same but with the signal recycling mirror transmissivity decreased from 0.2 to 0.02. For both graphs, the detunings, expressed in terms of phase shifts, are given b: A) 0 deg [tuned mode] B) 20 deg C) 25.2 deg D) 36 deg E) 54 deg*

The values presented in the table below were used in calculating the response.

**Table 1:** *Values used in plotting Fig. 5, taken from reference 16*

| | |
|---|---|
| $r_2 = .9999$; | $r_{AB} = \text{sqrt}(1-.014)$; |
| $k_c = 2*\pi/((1064*10^{\wedge}(-9)))$; | $t_{AB} = \text{sqrt}(.014)$; |
| $c = 3*10^{\wedge}8$; | $a = .991$; |
| $w = k_c * c$; | $m = 5.420675*10^{\wedge}7$; |
| $h = 10^{\wedge}(-12)$; | $L_{prc} = 2\pi m / k_c$ |
| $A = 1/25.65$; | $\phi_c = -\phi_{t1s+}\vert_{f_g=0}$ |
| $r_D = \text{sqrt}(1-.03)$; | $n = 3.75446*10^{\wedge}9$; |
| $t_D = \text{sqrt}(.03)$; | $L = (2\pi n + \phi_{r1c}/2)/k_c$ |
| $r_C = \text{sqrt}(1-.2)$; | $L_{srcSymMD0} = (2*\pi*(10.53157*10^{\wedge}7) + \pi)/(2*kc)$ |
| $t_C = \text{sqrt}(.2)$; | |

Both *h*, the amplitude of the GW, and *A*, the amplitude of the homodyning beam, are simple scale factors in these equations, appearing only as multiplicative constants. Their values are arbitrarily adjusted to normalize the response to one for the tuned case at zero GW frequency. The lower case "a" is the factor by which the field is assumed to be reduced on each pass through the SRC due to losses, and multiplies the reflectivity of the SRM in the Fabry-Perot calculations of the SRC reflectivity, ie $r_c \rightarrow a \times r_c$. The variables $r_2$, $r_{AB}$, $t_{AB}$, $r_C$, $t_C$, $r_D$, and $t_D$, represent the reflectivity and transmittivity of the mirrors labeled $M_2$, $M_{AB}$, $M_C$ and $M_D$, respectively, in Figs. 3 and 4. These values are taken from reference 13, and are the currently planned values for the Advanced LIGO system.

$L_{srcSymMD0}$ represents the symmetrically tuned ("mode 0") length of the SRC. The reflectivity of the SRC is calculated from standard Fabry-Perot theory starting from this value for the length of the cavity, with a variable detuning. In these calculations, the fact that one of the cavity mirrors has its substrate facing inwards must be taken into account. This alters the phase of the reflectivity of that mirror by 180 degrees, and thus alters the resonant length. The same is true of the PRC. Again, the SRC need not be 100% transmitting and is not, even on resonance, due to the mismatch in $r_{AB}$ and $r_C$. The more reflective the SRC is, the higher the signal will be, so long as the reflectivity of SRC still is small enough to allow the relevant sideband spectrum to fit within the bandwidth of the arms. The transmittivity of the SRC, though not unity, is nevertheless maximized for the chosen mirror reflectivities in these tuned mode plots. The reflectivity is higher, and the transmittivity lower, in detuned mode, but the magnitude cannot be chosen independently from the phase. This reflection phase could run between zero and $2\pi$ if the mirrors $r_{AB}$ and $r_C$ were matched, but this would mean lowering the reflectivity to zero in tuned mode, which would not be ideal. The attempt to resonate higher frequency sidebands in detuned mode therefore comes at the expense of the signal in tuned mode operation, and the mirror reflectivities $r_{AB}$ and $r_C$ are chosen with this tradeoff in mind.

Clearly hybrid modes of operation exist, with different choices for the lengths of the SRC and PRC and different choices for the phase of the carrier frequency beam with which the sidebands beat, but these two cases are enough to give a general idea of the behavior in the broadband tuned mode vs. narrowband detuned mode using the Advanced LIGO parameters.

## 4. Dispersive Effects

Having written the output of the device in terms of the sideband wavenumbers $k_+$ and $k_-$, we are now in a position to include easily the effects of dispersion on the system. The effect of the medium is to change the wavelength of light within it, so that $\lambda_{medium} = \lambda_{vacuum}/n$, where $n$ is the index of refraction of the medium. Equivalently, we may multiply the wavenumber by $n$, i.e. $k_{medium} = nk_{vacuum}$.

If we had a Fabry-Perot cavity of length $L$, the propagation phase light would ordinarily pick up on traveling from one end to the other is

(24) $\theta_{vacuum} = kL$ (where $k = k_{vacuum}$)

If, however, we assume a cavity of length $L$ partially filled by a medium of length $l$, that phase becomes

(25) $\theta = k(L-l) + n(\omega) k l$

The phase picked up by light making one round trip in the cavity is then

(26) $\theta_{r.t.} = 2k(L-l) + 2n(\omega) k l$

Assuming the light does not pick up any additional phase shifts as it propagates, the resonance condition is

(27) $\theta_{r.t.} = 2\pi m$ (*for integer m*)

If the light does pick up some phase shift, e.g. by reflecting off of a phase shifting mirror, then the resonance condition is altered so that the total phase picked up is equal to $2\pi m$, and the round trip propagation phase is equal to $2\pi m$ minus the extra phase due to the reflection.

In either case, resonance requires that the round trip phase $\theta_{r.t.}$ be equal to some predetermined constant. In free space, there would be only one value of $\omega$ which would fulfill the resonance condition. In a medium, however, we may have $\theta_{r.t.}$ depend on $\omega$ in a non-linear way. If we require that

(28) $\left. \dfrac{d\theta_{r.t.}}{d\omega} \right|_{\omega_0} = 0$

at some frequency $\omega_0$, then the round trip phase will not change with frequency at all for very small deviations from $\omega_0$, and will change by very small amounts for some range of frequencies around $\omega_0$. If $\omega_0$ happens to be the resonant frequency of the cavity, then a range of frequencies around $\omega_0$ will also be very close to resonance. The key to making a WLC is to make this range sufficiently large that the cavity resonates over a much wider bandwidth than it would if it were empty.

Substituting $k = \omega/c$ into equation 26 (since $k$ here is the vacuum wavenumber), and taking the derivative, we find

(29) $\left. \dfrac{d\theta_{r.t.}}{d\omega} \right|_{\omega_0} = \left[ \dfrac{d}{d\omega}\left( 2\dfrac{\omega}{c}(L-l) + 2n(\omega)\dfrac{\omega}{c}l \right) \right]_{\omega_0}$

$\approx 2\left( \dfrac{L}{c} + \left. \dfrac{dn}{d\omega}\right|_{\omega_0} \dfrac{\omega_0}{c} l \right)$ (*if* $n(\omega_0) \approx 1$)

Therefore the condition $\left. \dfrac{d\theta_{r.t.}}{d\omega} \right|_{\omega_0} = 0$ requires that

(30) $$\left.\frac{dn}{d\omega}\right|_{\omega_0} = \frac{-L}{l}\frac{1}{\omega_0}$$

The simplest model for a WLC assumes an index of refraction which is linear $\omega$ and has a slope given by the equation above:

(31) $$n(\omega) = 1 + \frac{-L}{l}\frac{1}{\omega_0}(\omega - \omega_0)$$

More complete models might assume $n(\omega)$ has the lineshape of the derivative of a Lorenztian, and choose the coefficients in the equation for this lineshape to give the appropriate slope at the center, or even more realistically, reproduce the lineshape of an index due to double gain peaks [2], for example, again with coefficients chosen such that the index has the appropriate slope between the two peaks.

Whichever functional form of $n(\omega)$ we choose, we may plug it into equation 25 to find its effect on the phase of light propagating through the cavity. The linear form of $n(\omega)$, for instance, gives

(32) $$\theta = k(L-l) + \frac{-L}{l}\frac{1}{\omega_0}(\omega - \omega_0)(kl)$$
$$= k(L-l) + \frac{-L}{k_0}(k - k_0)(k)$$

where $k_0 = \omega_0/c$ and $k$ is the vacuum wavenumber. All standard Fabry-Perot cavity analysis still applies, provided we use this expression for the propagation phase of light traveling from one end to the other of the cavity.

In general, in order to find the effect of changing the arm cavities into WLCs, on a LIGO type GW detector, we can make the following substitutions:

(33) $$k_+ L_S \rightarrow k_+(L_S - l) + n(k_+)k_+ l$$
$$k_- L_S \rightarrow k_-(L_S - l) + n(k_-)k_- l$$
$$k_c L_S \rightarrow k_c(L_S - l) + n(k_-)k_- l$$

Where $n(k)$ has the appropriate slope at the resonant frequency. Note that these expressions imply that we are placing the medium in the cavity of length $L_S$. In the type of system illustrated in Fig. 3, where $L_S = L$, this means the medium must be placed in the arms of the interferometer. In the case illustrated by Fig. 2, however, the medium may be placed between the beamsplitter and the Signal Recycling Mirror. In any case, we want to place it in whatever cavity stores the sidebands and has a length on the order of 4 km, in order to broaden the ordinarily very narrow linewidth associated with such a long resonator. Later in this section, we show a modified version of the configuration in fig. 3 where the WLC effect can be realized by placing the medium between the SRM and an auxiliary mirror.

In the expressions above, we have used the linear form of $n(\omega)$, without indicating a turn around point for the index function. Using more realistic functional forms of $n(\omega)$ gives a more realistic analysis of the behavior of the system. In Fig. 6, we plot the effects on the output of the interferometer using a slightly more complex model for the index, which assumes its lineshape is that of the derivative of a Lorentzian with a linewidth of approximately 16kHz, with the scaling of the Lorentzian function chosen to give an appropriate slope to the index function at the center. The bandwidth will, in practice, depend on the specific choice of dispersive material. In our experimental demonstrations of the WLC [10], we have seen a linear bandwidth approximately 5 MHz which is considerably broader than the GW bandwidth of interest. For these simulations, however, we have chosen to assume a narrower linewidth in order to show clearly the effect of the material on gravitational sideband frequencies near or outside of that linear bandwidth. For convenience we have chosen $l = L$ for these plots.

In general a WLC produces the same peak response as an empty cavity, but with a broader bandwidth. However, in the plots below the peak WLC response is twice as large as the peak of the detuned mode response. This is because in the WLC gravitational wave detector, both GW sidebands resonate instead of just one.

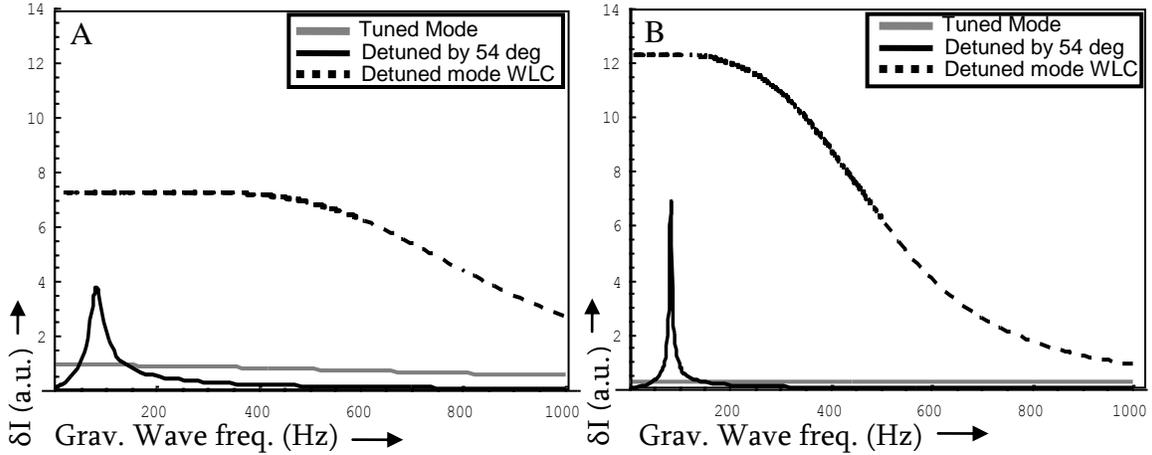

**Figure 6:** *A) Tuned mode response, detuned mode response, and detuned mode response with WLC for a cavity/interferometer with Advanced LIGO reflectivities and lengths. B) The same, but using an SRM with a transmissivity of $t_c^2=0.02$ instead of $t_c^2=0.2$ as in Advanced LIGO. In both graphs the dispersive material is chosen to have $dn/d\omega = -L/(\omega_0 l)$ over a linewidth of approximately 1600 Hz centered around that resonant frequency, but the output begins to fall when the other sideband is no longer within the linear region.*

These graphs illustrate that the WLC-based GW detector isn't just broader in bandwidth than the currently planned Advanced LIGO model, but potentiallymore sensitive as well, since the mirror reflectivities could be optimized for sensitivity in narrowband mode rather than for bandwidth.

The cases illustrated above are for the Advance LIGO type detector illustrated in Fig. 3. We have also carried out the derivation, and developed the frequency response graph, for the type of detector illustrated in Fig. 2, with and without a WLC. This system is similar to that for which we have already plotted the output, except that the finesse of the arm cavities for the SBs cannot be dynamically controlled – it is given by the reflectivity of the SRM and does not vary with its position. Nevertheless, the behavior in detuned mode is very similar. Incorporating a WLC again gives a broadband response equal to twice the peak value of the narrowband response, since with the WLC, both sidebands will resonate along with the carrier. This is illustrated in figure 7. In this case, it is worth noting that again, increasing the reflectivity of the SRM increases the sensitivity of the detector without compromising the bandwidth, so that with the incorporation of a WLC, a detector of this type can be made far more sensitive, for a given carrier power, than would be feasible without the WLC.

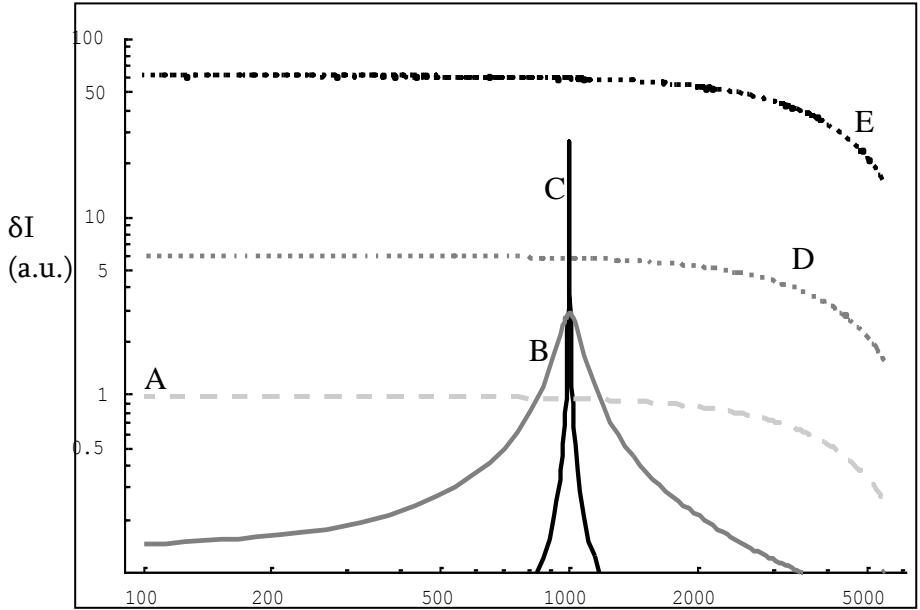

**Figure 7:** Illustration of the effect of WLC on the configuration shown in fig.2: A) $r_{SRM} = 0$, B) Detuned mode, $r_{SRM} = \sqrt{1-0.1}$, C) Detuned mode, $r_{SRM} = \sqrt{1-0.001}$
D) Same as B but with WLC E) Same as C but with WLC

As discussed in reference 8, this simpler type of design may in fact be more suitable for use with a WLC than the design illustrated by Fig. 3. In this configuration, illustrated by Fig. 2, the dispersive medium may be placed outside of the interferometer arms, between the beamsplitter and the signal recycling mirror. In this case, the beam power within the medium can be smaller, since the high power carrier is not incident on the medium in this configuration. The design illustrated by figure 2 also requires a simpler control system, with fewer cavities to lock. It does suffer from the issues regrading the heating of the beamsplitter as described earlier, and is slightly less flexible.

The two WLC-enabled designs considered above each suffers from a significant practical constraint. In the first case, the WLC element is added inside the arm cavities of fig. 3. Since this requires two separate WLC elements to be inserted in the two arms, it would be difficult to match dispersion exactly. Furthermore, since the beams inside the arm cavities are very large in diameter, the WLC elements have to be very large as well. The best WLC medium we have identified for the wavelength used in LIGO apparatus is a photorefractive crystal[3]. It would be very difficult to make such a crystal big enough to meet this requirement. Finally, the presence of the very strong pump beams would most likely cause heating problems in the WLC medium. In the second design, the WLC element is placed between the beamsplitter and the SRM of the configuration shown in fig. 2. However, the configuration of fig. 2 has already been discounted in the LIGO as well as the AdLIGO design because of the problem of heating of the beamsplitter.

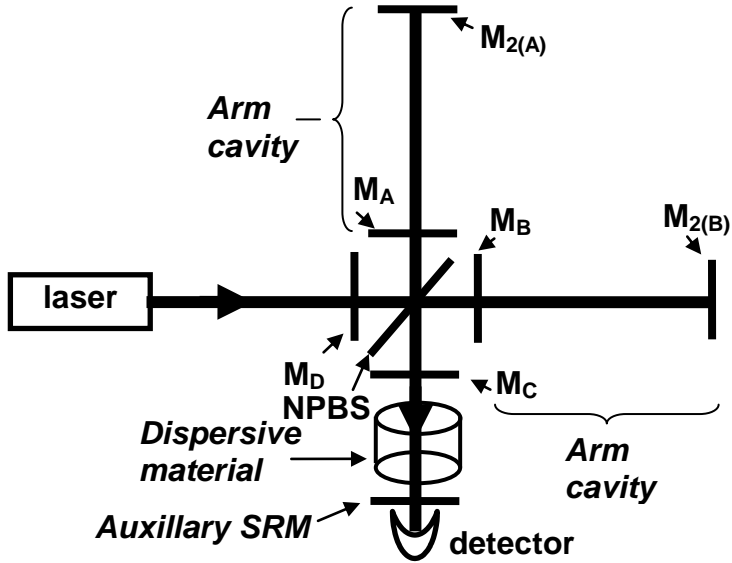

**Figure 8:** Dual recycling system with auxillary signal recycling mirror, allows for large carrier build-up in arms, low power on primary beamsplitter, and, if the reflectivity of $M_C$ is matched to that of $M_{AB}$, for high finesse signal recycling over broad band using WLC.

In figure 8, we offer a new WLC-enabled design that circumvents all these practical constraints. In this design $M_{AB}$ and $M_C$ have matched reflectivities. If the distance between them is such that the carrier is resonant, and the linewidth broad enough that the relevant range of sidebands resonate as well, these two mirrors effectively disappear for signal recycling. The cavity they form is transparent to the sideband light and causes no effective phase shift. The sidebands then encounter the auxilliary mirror, in front of the detector, and are reflected back through the transparent cavity again into the arms. This system is, as far as the sidebands are concerned, the same as that modeled in Fig. 2. However, it allows the power in the arms to be kept high while keeping the power on the beamsplitter low. Finally, it allows us to place the dispersive medium outside of the interferometer arms, between the two mirrors which lie between the beamsplitter and the detector. The design proposed in fig. 8 can be implemented in a non-invasive manner by a minor modification of the current AdLIGO design. The beam size before the auxiliary mirror and after the SRM can be reduced using lenses in order to accommodate the size of a photorefractive crystal.

We simulated this situation with $r_C = r_{AB} = \sqrt{1-0.014}$ for two different values of the auxillary mirror reflectivity. The distance between $M_{AB}$ and $M_C$ was reduced to 0.5 m, while the distance between $M_C$ and the auxilliary mirror was chosen to be ~057m. The results, with and without the anomalously dispersive material, are shown below.

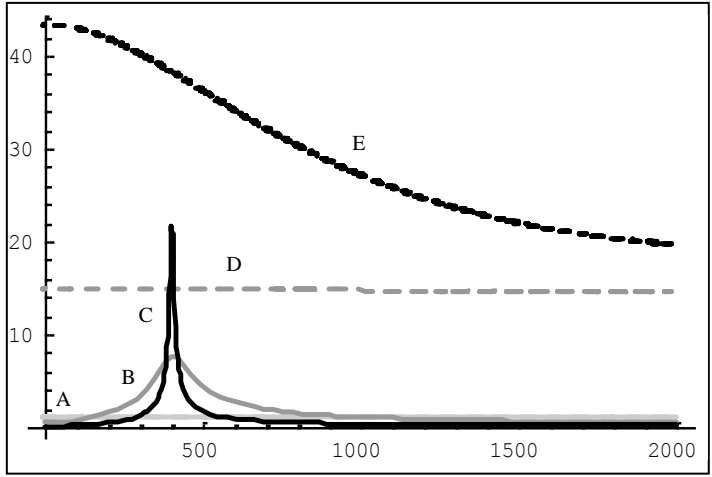

**Figure 9:** Response functions for the system depicted in Figure 8. A) $r_{AUX} = 0$, no dispersive material B) $r_{AUX} = \sqrt{1-0.02}$, no dispersive material. C) $r_{AUX} = \sqrt{1-0.002}$, no dispersive material D) Same as B but including material with critically anomalous dispersion at location shown in Figure 8. E) Same as C but including material with critically anomalous dispersion at location shown in Figure 8.

## 5. Conclusion

Almost all Michelson-based GW detectors can be described by equation 22 of this document. This equation follows from basic Fabry-Perot theory, provided the arms of the interferometer are identical, and involves treating some pairs of mirrors as a single compound mirror with a frequency dependent reflectivity, in certain cases.

The effect of introducing a medium into such a system is to change the propagation phase of the light within the long arm cavities. If the medium has a negative dispersion with a slope given by equation 30, that propagation phase will not vary with frequency over some range, and the resonance bandwidth of the cavity will be broadened. This will have the effect, for all of the variations on the Michelson interferometer that we have discussed, of both broadening the bandwidth of the detector and increasing its sensitivity to some degree by preventing destructive interference within the cavity.

The method given here for calculating the effect of a WLC on any Michelson-based GW detector allows us to consider a variety of different designs, each with its own advantages and disadvantages. The graph shown in Fig. 9 shows that a WLC could significantly improve a GW detector, with enhanced sensitivity as well as bandwidth, building on the design currently proposed for Advanced LIGO.

---